\documentclass[twocolumn]{IEEEtran}

\usepackage{cite}

\ifCLASSINFOpdf
  \usepackage[pdftex]{graphicx}
\else
\fi
\usepackage{amsmath}
\usepackage{subfigure}
\usepackage{algorithm}
\usepackage{algpseudocode}

\begin{document}
%
\title{Faster-than-Nyquist Non-Orthogonal Frequency-Division Multiplexing for Visible Light Communications}
%
%
%

\author{Ji~Zhou,~
        Yaojun Qiao,~
        Qi Wang,~
        Jinlong Wei,~
        Qixiang Cheng,~
        Tiantian Zhang,~
        Zhanyu Yang,~
        Aiying Yang,~
        and Yueming Lu
\thanks{This work was supported in part by National Natural Science Foundation of China (61475094, 61331010); National Key Research and Development Program (2016YFB0800302); BUPT Excellent Ph.D. Students Foundation; China Scholarship Council Foundation.~\emph{(Corresponding author: Yaojun Qiao.)}}
\thanks{Ji~Zhou, Yaojun~Qiao, and Tiantian Zhang are with the State Key Laboratory of Information Photonics and Optical Communications, School of Information and Communication Engineering, Beijing University of Posts and Telecommunications (BUPT), Beijing 100876, China (e-mail: zhouji@bupt.edu.cn and qiao@bupt.edu.cn).}
\thanks{Ji~Zhou and Qixiang Cheng are with the Department of Electrical Engineering, Columbia University, New York 10026, USA.}
\thanks{Qi Wang is with Electronics and Computer Science, University of Southampton, Southampton SO17 1BJ, U.K.}
\thanks{Jinlong Wei is with the with the Huawei D$\ddot{\text{u}}$sseldorf GmbH, European Research Center, M$\ddot{\text{u}}$nchen 80992, Germany}
\thanks{Zhanyu Yang is with the Department of Electrical and Computer Engineering, University of Virginia, 351 McCormick Road, Charlottesville, Virginia 22904, USA.}
\thanks{Aiying Yang is with the School of Optoelectronics, Beijing Institute of Technology, Beijing 100081, China. }
\thanks{Yueming Lu is with the Key Laboratory of Trustworthy Distributed Computing and Service, Ministry of Education, Beijing University of Posts and Telecommunications (BUPT), Beijing 100876, China.}
}

\renewcommand{\arraystretch}{1.2}

\maketitle
\pagestyle{empty}  
\thispagestyle{empty} 

\begin{abstract}
\boldmath
In this paper, we propose a faster-than-Nyquist (FTN) non-orthogonal frequency-division multiplexing (NOFDM) scheme for visible light communications (VLC) where the multiplexing/demultiplexing employs the inverse fractional cosine transform (IFrCT)/FrCT. Different to the common fractional Fourier transform-based NOFDM (FrFT-NOFDM) signal, FrCT-based NOFDM (FrCT-NOFDM) signal is real-valued, which can be directly applied to the VLC systems without the expensive upconversion and thus it is more suitable for the cost-sensitive VLC systems. Under the same transmission rate, FrCT-NOFDM signal occupies smaller bandwidth compared to OFDM signal. When the bandwidth compression factor $\alpha$ is set to $0.8$, $20\%$ bandwidth saving can be obtained. Therefore, FrCT-NOFDM has higher spectral efficiency and suffers less high-frequency distortion compared to OFDM, which benefits the bandwidth-limited VLC systems. As the simulation results show, bit error rate performance of FrCT-NOFDM with $\alpha$ of $0.9$ or $0.8$ is better than that of OFDM. Meanwhile, FrCT-NOFDM has a superior security performance. In conclusion, FrCT-NOFDM shows the potential for application in the future VLC systems.
\end{abstract}


%
\IEEEpeerreviewmaketitle

\section{Introduction}
\IEEEPARstart{R}{ecently}, visible light communications (VLC) have been proposed to provide high-speed network access for office, shop center, warehouse, and airplane because of many advantages such as the lost-cost front-ends, unregulated huge frequency resources, and no electromagnetic interference \cite{Biagi1, Elgala1, Gross1,Nan1}. As a potential access option for the future 5G wireless systems, VLC systems are gaining extensive attention \cite{Wu1, Rahaim1, Ayyash1}. However, there are many obstacles for the practical VLC systems. For instance, the data rate of VLC systems is limited by the bandwidth of the light-emitting diodes (LEDs) and multipath effect \cite{Ghassemlooy1, Chi1, Biagi2}. The multipath effect gives rise to the frequency-selective power fading, which seriously limits the effective bandwidth. The longer the delay spread of multipath is, the smaller the effective bandwidth is. How to transmit more data on the limited bandwidth is an important issue in VLC systems.

In order to fully utilize the limited bandwidth of VLC systems, the modulation schemes with high spectral efficiency should be employed. Orthogonal frequency-division multiplexing (OFDM) is a well-known modulation scheme with high spectral efficiency, which has been widely investigated for VLC systems \cite{Tsonev1, Elgala2, Wang1, Wang2, Mossaad1}. As we know, VLC is an intensity-modulation/direct-detection (IM/DD) optical system. Therefore, OFDM signal for the VLC systems needs to be real-valued and unipolar. For widely-used discrete Fourier transform-based OFDM (DFT-OFDM), Hermitian symmetry should be used to generate the real-valued signal. To obtain the unipolar signal, there are two popular approaches: one is DC-biased optical OFDM (DCO-OFDM) and the other is asymmetrical clipping optical OFDM (ACO-OFDM) \cite{Armstrong1, Armstrong2, Zhou1}. In DCO-OFDM, the biasing and clipping operations are employed to make the bipolar OFDM signal unipolar. In ACO-OFDM, only odd subcarriers are utilized to transmit data, whereby the negative part of the bipolar OFDM signal is redundant. Therefore, the bipolar OFDM signal can be converted into an unipolar signal by clipping the negative part. DCO-OFDM and ACO-OFDM have their respective advantages and disadvantages and thus we can select the suitable one depending on the application scenario.

\begin{figure*}[!t]
\centering
\includegraphics[width=6.4in]{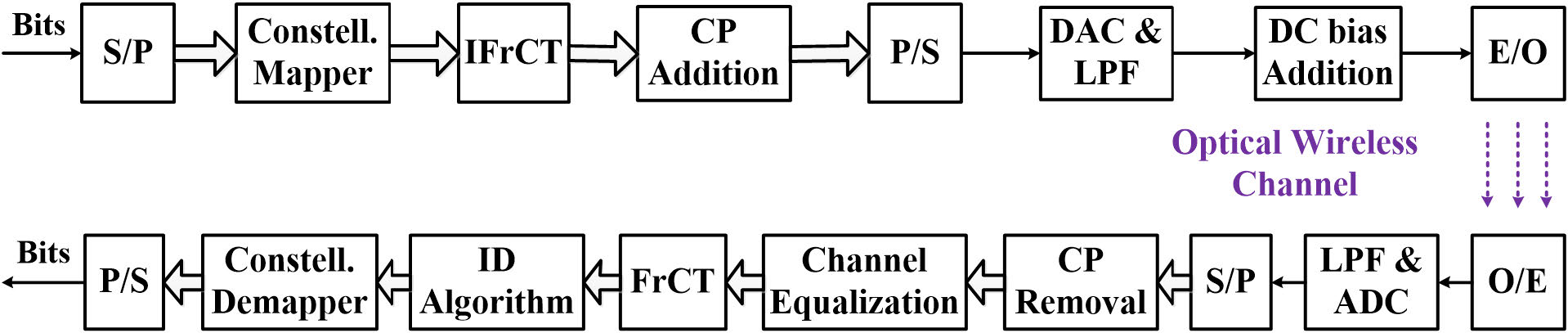}
\caption{Block diagram of FTN FrCT-NOFDM for VLC systems. S/P: series-to-parallel, P/S: parallel-to-series, IFrCT: inverse fractional cosine transform, FrCT: fractional cosine transform, CP: cyclic prefix, DAC: digital-to-analog converter, ADC: analog-to-digital converter, LPF: low-pass filter, E/O: electro/optic, O/E: optic/electro, ID: iterative detection.}\label{fig:1}
\end{figure*}

Recently, non-orthogonal frequency-division multiplexing (NOFDM) systems have been proposed in both wireless and optical communications to further improve the spectral efficiency by compressing the subcarrier spacing \cite{Darwazeh1, Y1, Y2, Xu1, Nopchinda1, Zhou2, Zhou3}. In conventional NOFDM for IM/DD optical systems, the multiplexing usually employs inverse fractional Fourier transform (IFrFT)\cite{Darwazeh1, Y1}. However, the subcarrier distribution in FrFT-based NOFDM (FrFT-NOFDM) is different from that in OFDM, thus the real-valued signal cannot be generated by employing Hermitian symmetry\cite{Zhou2}. In general, the upconversion is applied to the complex-valued NOFDM signal for realizing intensity modulation\cite{Darwazeh1, Y1}. There are two methods to implement the upconversion: analog and digital methods. In the analog method\cite{Darwazeh1}, the digital complex-valued NOFDM signal requires two-channel digital-to-analog converters (DACs) to generate the analog signal. Then, an in-phase/quadrature (I/Q) mixer with a radio frequency (RF) waveform is required to combine two-channel analog signals and up-convert to intermediate frequency. The two DACs, I/Q mixer, and RF waveform are uneconomical for the VLC systems. In the digital method\cite{Y1}, the upconversion can be implemented by the digital signal processing, which can avoid some expensive analog devices but requires the high-bandwidth DAC. In both analog and digital methods, the generated intermediate-frequency signal has higher bandwidth compared to the baseband signal, which requires the electrical, electro-optic and photoelectric devices with high electrical bandwidth. The high-bandwidth devices are expensive for the cost-sensitive VLC systems.

In our previous work, we proposed the real-valued faster-than-Nyquist (FTN) NOFDM schemes, which have been experimentally demonstrated in the fiber-optic communications \cite{Zhou2, Zhou3}. FTN signal was first proposed by Mazo in 1975, which can achieve the symbol rate faster than the Nyquist rate\cite{Mazo}. For the IM/DD optical systems, the real-valued FTN NOFDM signal requires no upconversion and has high spectral efficiency. Under the same transmission rate, the real-valued FTN NOFDM signal occupies smaller bandwidth compared to OFDM signal, which is promising in bandwidth-limited VLC systems. In this paper, we apply FTN NOFDM signal to VLC systems and study its performance in detail.

The main contributions of this paper are as follows:
\begin{itemize}
  \item  A novel FTN fractional cosine transform-based NOFDM (FrCT-NOFDM) signal for VLC systems: The bandwidth of FrCT-NOFDM signal is smaller than that of OFDM signal thereby having higher spectral efficiency. Meanwhile, FrCT-NOFDM signal is real-valued. Therefore, it is suitable for the bandwidth-limited and cost-sensitive VLC systems.
  \item  A statistical characteristic of FrCT-NOFDM signal: We verify that FrCT-NOFDM samples can be approximated as independent identically distributed (i.i.d.) Gaussian random variables. Under this condition, the statistical characteristic of DC-biased FrCT-NOFDM is derived.
   \item  A detailed analysis for the performance of FrCT-NOFDM on the optical-wireless channel: The bit error rate (BER) performance of FrCT-NOFDM is comprehensively analyzed under the different bandwidth compression factor, root-mean-square (RMS) delay spread, DC bias, and iterative number. Meanwhile, we demonstrate that FrCT-NOFDM has the superior security performance.
\end{itemize}

The rest of this paper is organized as follows. We give the principle of FTN FrCT-NOFDM for the VLC systems in Section \ref{section_2}. In Section \ref{section_3}, we demonstrate the statistical characteristic of FrCT-NOFDM. In Section \ref{section_4}, we analyze the optical-wireless channel and noise models for simulation. In Section \ref{section_5}, we implement the simulations and give the simulation results for studying the performance of FrCT-NOFDM in the VLC systems. Finally, the paper is concluded in Section \ref{section_6}.

\section{Principle of FTN FrCT-NOFDM for VLC Systems}\label{section_2}
A block diagram of FTN FrCT-NOFDM for VLC systems is depicted in Fig. \ref{fig:1}. Different from the FrFT-NOFDM, the inverse FrCT (IFrCT)/FrCT algorithm is employed to realize the multiplexing/demultiplexing processing in FrCT-NOFDM. The $N$-order IFrCT and FrCT are defined as
\begin{equation}\label{eq:eq1}
x_{n}=\sqrt{\frac{2}{N}}\sum_{k=0}^{N-1}W_{k}X_{k}\text{cos}(\frac{\pi \alpha (2n+1)k}{2N}),
\end{equation}
\begin{equation}\label{eq:eq2}
X_{k}=\sqrt{\frac{2}{N}}W_{k}\sum_{n=0}^{N-1}x_{n}\text{cos}(\frac{\pi \alpha (2n+1)k}{2N})
\end{equation}
where $0\leq n\leq N-1$, $0\leq k\leq N-1$,
\begin{equation}\label{eq:eq3}
W_{k} =
\begin{cases}
\frac{1}{\sqrt{2}}, &\text{$k=0$}\\
1, &\text{$k=1,~2,~\cdots,~N-1$}
\end{cases}
\end{equation}
and $\alpha$ is the bandwidth compression factor. The $\alpha$ less than $1$ determines the level of the bandwidth compression. IFrCT is a real-valued transform, thus if the input of IFrCT is a real-valued constellation such as $M$ pulse-amplitude modulation ($M$-PAM), the output is a real-valued NOFDM signal. Therefore, FrCT-NOFDM needs no Hermitian symmetry to generate the real-valued signal. For IM/DD optical system, the upconversion is not required in FrCT-NOFDM signal.

\begin{figure}[!t]
\centering
\includegraphics[width=3.4in]{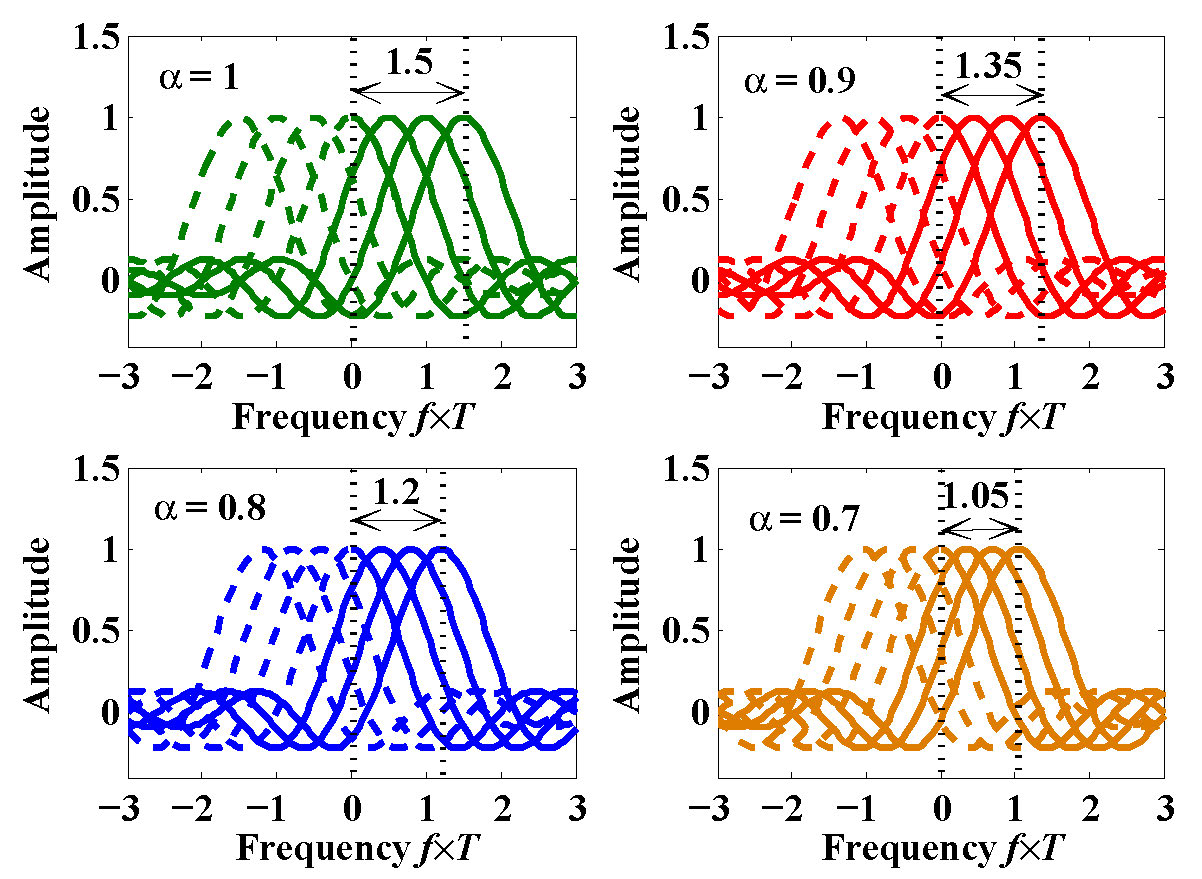}
\caption{Sketched spectra of DCT-OFDM (i.e., $\alpha = 1$) and FrCT-NOFDM when the number of subcarriers is set to $4$.}\label{fig:2}
\end{figure}

As a special case, Eq. (\ref{eq:eq2}) is the Type-II discrete cosine transform (DCT) in which the matrix is orthogonal when $\alpha$ is set to $1$. The Type-II DCT is probably the most commonly used form, which is often simply referred to as ``the DCT'' \cite{Ahmed1}. DCT-based OFDM (DCT-OFDM) signal for IM/DD optical systems can be generated by Type-II DCT and has the same BER performance compared to frequently-used DFT-OFDM signal\cite{Ouyang, Zhou4, Zhou5}.

Figure \ref{fig:2} shows the sketched spectra of DCT-OFDM (i.e., $\alpha = 1$) and FrCT-NOFDM when the number of subcarriers is set to $4$. The subcarrier spacing of DCT-OFDM is equal to $1/2T$ where $T$ denotes the time duration of one DCT-OFDM symbol \cite{Tan1, Zhao1, Giacoumidis1, Giacoumidis2}. In DCT-OFDM, all the subcarriers locate on the positive frequency region and their mirror images fall on the negative frequency region. Due to the compression of subcarrier spacing, the subcarrier spacing of FrCT-NOFDM is equal to $\alpha/2T$. When the number of subcarriers is large enough, the bandwidth of FrCT-NOFDM can be defined as
\begin{equation}\label{eq:eq4}
B = \frac{\alpha (N-1)}{2T}+\frac{1}{T} \approx \frac{\alpha N}{2T}.
\end{equation}
The bandwidth of FrCT-NOFDM decreases with the decrease of the $\alpha$. When $\alpha$ is set to 0.8, $20\%$ bandwidth saving is obtained. Therefore, FrCT-NOFDM has higher spectral efficiency compared to DCT-OFDM. As we know, the Nyquist rate of FrCT-NOFDM is equal to $2B$,
\begin{equation}\label{eq:eq5}
R_{N} = \frac{\alpha N}{T}.
\end{equation}
The symbol rate of FrCT-NOFDM is equal to
\begin{equation}\label{eq:eq6}
R_{S} = \frac{N}{T}.
\end{equation}
Therefore, when the $\alpha$ is set to $0.8$, the symbol rate of FrCT-NOFDM is $25\%$ faster than the corresponding Nyquist rate.

\section{Statistical Characteristic of FrCT-NOFDM Signal}\label{section_3}
As we know, DCT-OFDM signal has the approximately Gaussian distribution \cite{Zhou4}. What is the distribution of FrCT-NOFDM with the compressed subcarrier spacing? In this section, we give analysis of the statistical characteristic for the FrCT-NOFDM signal.

\begin{figure}[!t]
\centering
\subfigure[~~~~Probability density function]
{\label{fig_3:a}
\includegraphics[width=2.8in]{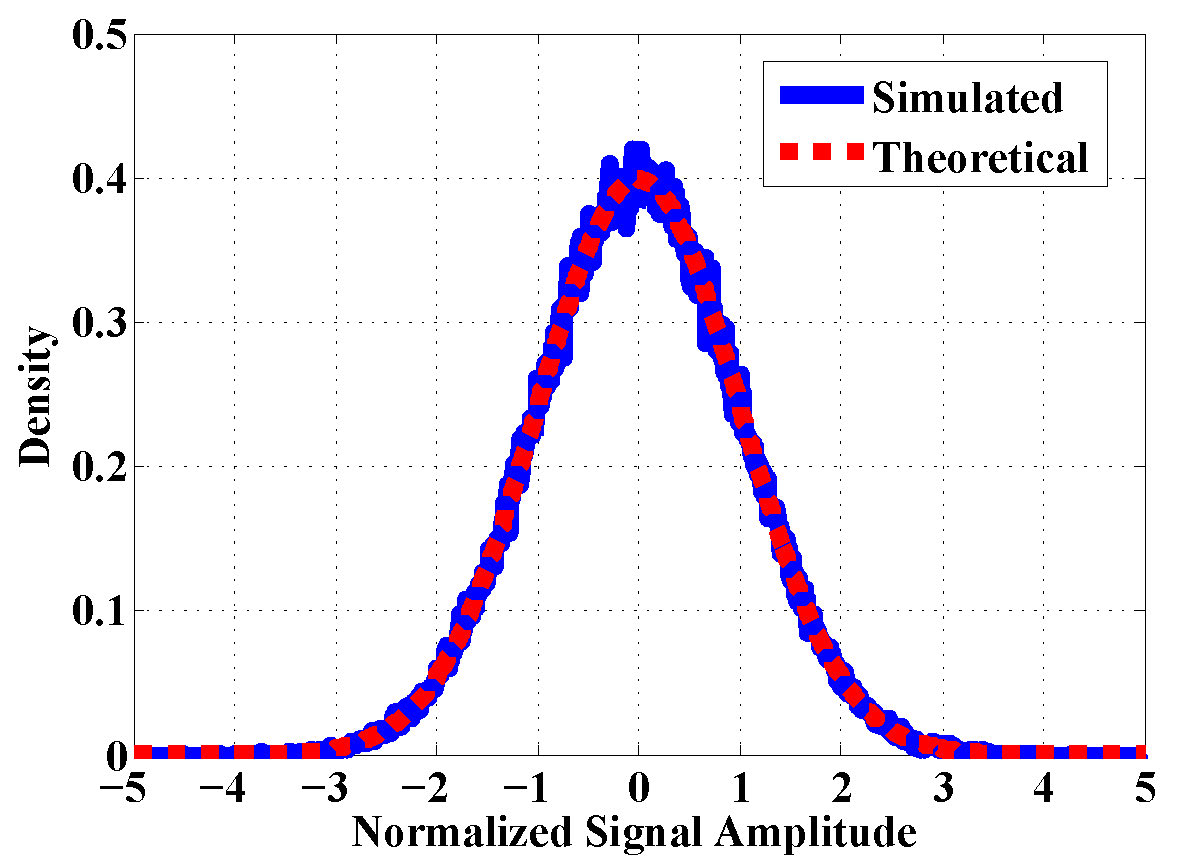}}
\subfigure[~~~~Cumulative distribution function]{
\label{fig_3:b}
\includegraphics[width=2.8in]{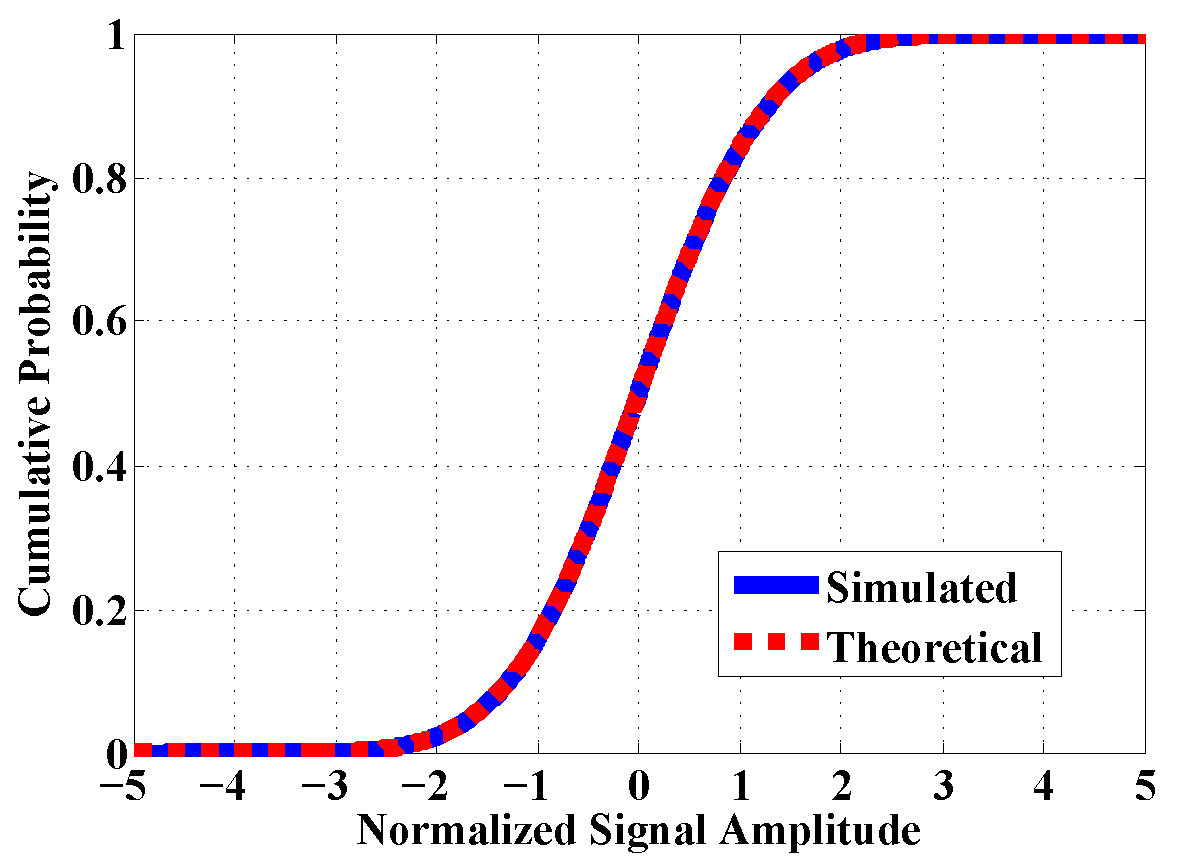}}
\caption{Approximately Gaussian distributed time-domain FrCT-NOFDM signal. FrCT-NOFDM signal is normalized such that its variance is equal to one.} \label{fig:3}
\end{figure}

The dash lines in Fig. \ref{fig:3} depict the theoretical results of probability density function and cumulative distribution function of the standard normal distribution. As the solid lines in Fig. \ref{fig:3} show, the statistical characteristic of FrCT-NOFDM signal is simulated when $\alpha$ is set to $0.8$ and the size of FrCT is set to 256. FrCT-NOFDM signal is normalized such that its variance is equal to one. Obviously, the simulated results of probability density function and cumulative distribution function agree well with the theoretical results. Therefore, we can model the FrCT-NOFDM signal $x_n$ using i.i.d. Gaussian process with the following probability density function,
\begin{equation}\label{eq:eq7}
pdf(x) = N(x;0,E[|x_n|^2])
\end{equation}
 with
\begin{equation}\label{eq:eq8}
N(x;\mu, \sigma^2) = \frac{1}{\sqrt{2\pi}\sigma}e^{-\frac{(x-\mu)^2}{2\sigma^2}}
\end{equation}
where $\mu$ and $\sigma^2$ are the mean and variance, respectively, of the Gaussian distribution.

For VLC systems, the transmitted signal should be real-valued and unipolar. FrCT-NOFDM signal is real-valued but bipolar. A biasing and clipping operations can be used to make the bipolar FrCT-NOFDM signal unipolar. Without loss of generality, we employ the single-sided clipping operation in VLC systems. The generated DC-biased FrCT-NOFDM signal can be defined as
\begin{equation}\label{eq:eq9}
s_{n} =
\begin{cases}
x_{n}+B_{DC},& \text{$x_{n}\geq -B_{DC}$}\\
0,& \text{$x_{n}<-B_{DC}$}
\end{cases}
\end{equation}
where $B_{DC}$ is DC bias, which is related to the power of $x_{n}$,
\begin{equation}\label{eq:eq10}
B_{DC} = k\sqrt{E\{x_n^2\}}.
\end{equation}
The size of DC bias can be defined as the power ratio of $s_{n}$ to $x_{n}$ in dB, i.e. $10\times\text{log}_{10}(k^2+1)$ dB.

The probability density function of the DC-biased FrCT-NOFDM signal $s_{n}$ is defined as
\begin{equation}\label{eq:eq11}
pdf_{s}(x) =
\begin{cases}
N(x;~B_{DC},~\sigma^2),& \text{$x \geq 0$}\\
Q(\frac{B_{DC}}{\sigma})\delta(x),& \text{$x=0$}
\end{cases}
\end{equation}
where $\delta(x)$ is the Dirac delta function with an unit impulse at $x$ only and the well-known $Q$ function is defined as the integral over the probability density function of the standard normal distribution,
\begin{equation}\label{eq:eq12}
Q(\upsilon) = \int_{\upsilon}^{\infty}N(\tau;~0,~1)d\tau = 1-\Phi(\upsilon).
\end{equation}

In general, the electrical power $P_{e}$ is proportional to $E[s_{n}^2]$. Without loss of generality, we define $P_{e}=E[s_{n}^2]$. Thus, the electrical power $P_{e}$ of DC-biased FrCT-FOFDM is given by
\begin{equation}\label{eq:eq13}
\begin{aligned}
P_{e}&=E[s_{n}^{2}]=\int_{-\infty}^{\infty}x^{2}\cdot pdf_{s}(x)dx\\
&=\sigma^{2}\Phi(\frac{B_{DC}}{\sigma})+B_{DC}^{2}\Phi(\frac{B_{DC}}{\sigma})+\frac{\sigma B_{DC}}{\sqrt{2\pi}}e^{-\frac{B_{DC}^2}{2\sigma^{2}}}.
\end{aligned}
\end{equation}
Therefore, the electrical power of DC-biased FrCT-NOFDM signal contains three parts: the power of the useful signal, DC bias, and clipping noise. The clipping operation induces the distortions including the attenuation of $x_n$ and clipping noise. Therefore, the useful signal can be written as $\eta \times x_{n}$ where $\eta$ is the attenuation factor, which is equal to $\sqrt{\Phi(\frac{B_{DC}}{\sigma})}$. With the increase of $B_{DC}$, the attenuation $\eta$ approaches to $1$ and the power of clipping noise decreases to $0$, but the power of DC bias increases. Therefore, when $B_{DC}$ is sufficiently large, Eq. (\ref{eq:eq13}) is simplified to
\begin{equation}\label{eq:eq14}
P_{e}=\sigma^{2}+B_{DC}^{2}.
\end{equation}

A large $B_{DC}$ can be used to eliminate the clipping-induced distortion, but it is inefficient in terms of the signal power because $B_{DC}$ cannot carry any information. For application in VLC systems, we need to employ a suitable $B_{DC}$ to make bipolar FrCT-NOFDM to be unipolar.

\section{Optical-Wireless Channel Model and Noise Model}\label{section_4}
The optical-wireless channel can be modeled as a linear baseband system. The received signal can be given as
\begin{equation}\label{eq:eq15}
r(t) = h(t)\ast x(t)+n(t)
\end{equation}
where $h(t)$ is the channel impulse response, $x(t)$ is the transmitted signal, $n(t)$ is the noise component, and $\ast$ denotes the convolution operation. The channel impulse response $h(t)$ can be defined as
\begin{equation}\label{eq:eq16}
h(t) = \sum^{N_{D}}_{n=0} h_{n}\delta(t-n\Delta\tau)
\end{equation}
where $N_D$ is the number of paths, $h_{n}$ is the channel coefficient and $n\Delta\tau$ is the delay of the $n$-th path.

\begin{figure}[!t]
\centering
\includegraphics[width=3in]{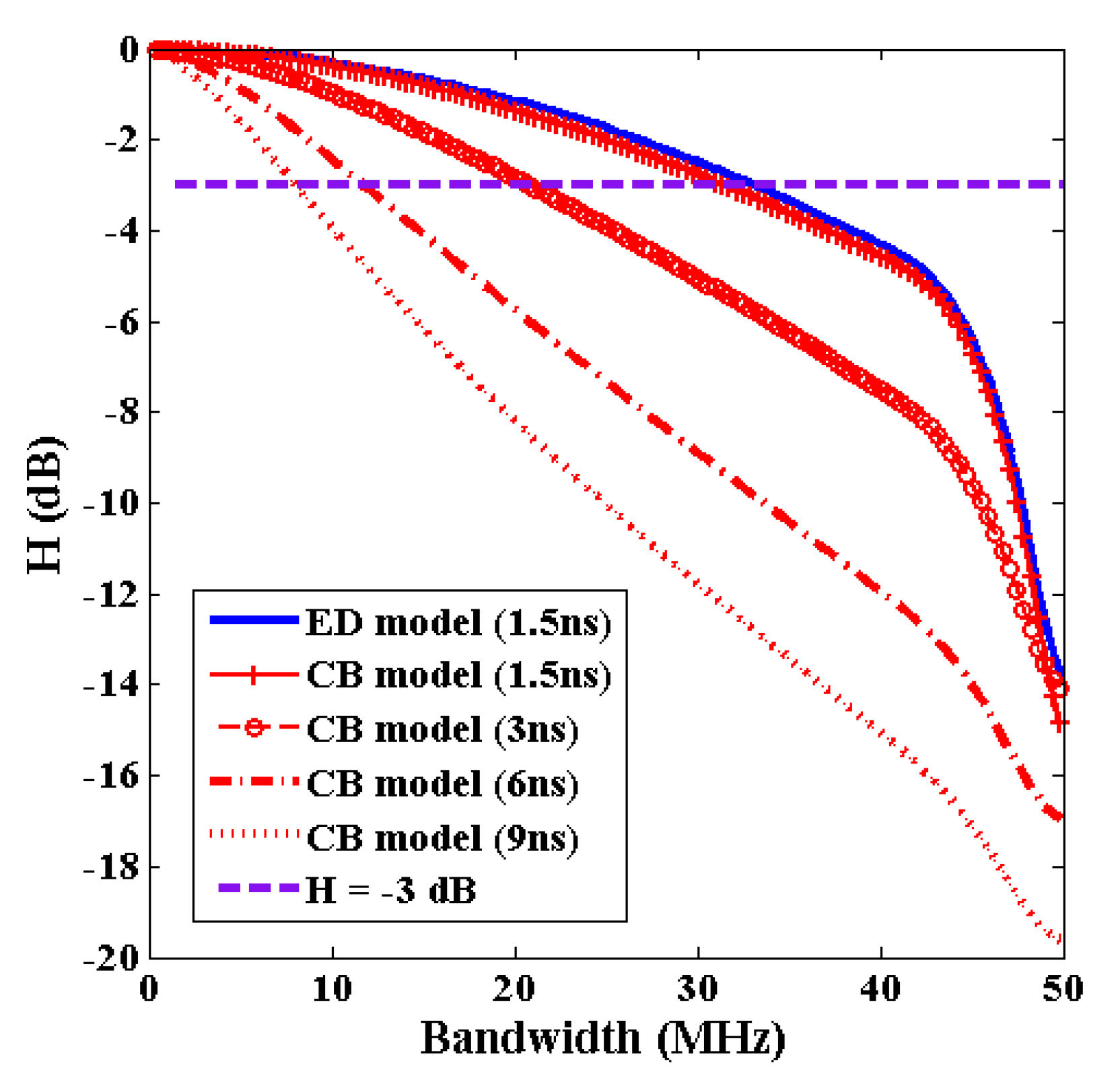}
\caption{The transfer function of the exponential-decay (ED) and ceiling-bounce (CB) models.}\label{fig:4}
\end{figure}

In general, the optical-wireless channel models fall into two categories: directed and non-directed (i.e. diffused) models \cite{Carruthers1, Kahn1, Jungnickel1}. In the directed model, line of sight (LOS) plays the major role. The directed model can be appropriately considered as an additive white Gaussian noise (AWGN) channel model. In the non-directed model, the optical power propagates along various paths with different lengths, which causes the multipath effect. In this paper, we employ the non-directed model to investigate the influence of the multipath effect on FrCT-NOFDM. The multipath effect is usually caused by two ways: one is the multiple-reflection light and the other is the single-reflection light. The exponential-decay (ED) and ceiling-bounce (CB) models were proposed to model both multiple reflection light and single reflection light \cite{Carruthers1}. The channel impulse response of the ED model can be defined as
\begin{equation}\label{eq:eq17}
h_{ed}(t) = \frac{1}{2D}e^{-\frac{t}{2D}}u(t)
\end{equation}
where the $D$ is the RMS delay spread of the multiple reflections and $u(t)$ is the unit step function. The channel impulse response of the CB model is given by
\begin{equation}\label{eq:eq18}
h_{cb}(t) = \frac{6a^6}{(t+a)^7}u(t)
\end{equation}
where $a=12\sqrt{\frac{11}{13}}D$.

Figure \ref{fig:4} shows the transfer function of the ED and CB models. When the RMS delay spread is set to $1.5$ ns, the $3$-dB bandwidth of ED model and CB model is $33.7$ MHz and $31.4$ MHz, respectively. Therefore, the 3-dB bandwidth of the ED model is slightly wider than that of the CB model for the same RMS delay spread. The $3$-dB bandwidth of the CB model is $20.9$ MHz, $11.7$ MHz, and $8.3$ MHz when the RMS delay spread is set to $3$ ns, $6$ ns, and $9$ ns, respectively. The 3-dB bandwidth of the CB model decreases with the increase of the RMS delay spread. The multipath effect causes the frequency-selective power fading, which seriously limits the effective bandwidth. For VLC systems, there are many other complex channel models, which are derived from the non-directed model by considering many other conditions such as the position of LED and photodiode (PD) and the field of view of the PD \cite{Chen1, Chvojka4, Lee1}. In this paper, we aim to investigate the performance of FrCT-NOFDM signal influenced by the multipath effect. Without loss of generality, we can employ the CB model to achieve the aim in the simulation.

In optical-wireless systems, there are two main noise components: receiver circuit thermal noise and photon noise, which are both independent of the transmitted signal and modeled as white Gaussian distribution \cite{ Kahn1, Jungnickel1}. Therefore, we can model the total noise $n(t)$ as the Gaussian and signal-independent distribution in the simulation.

\begin{figure}[!t]
\centering
\includegraphics[width=3.4in]{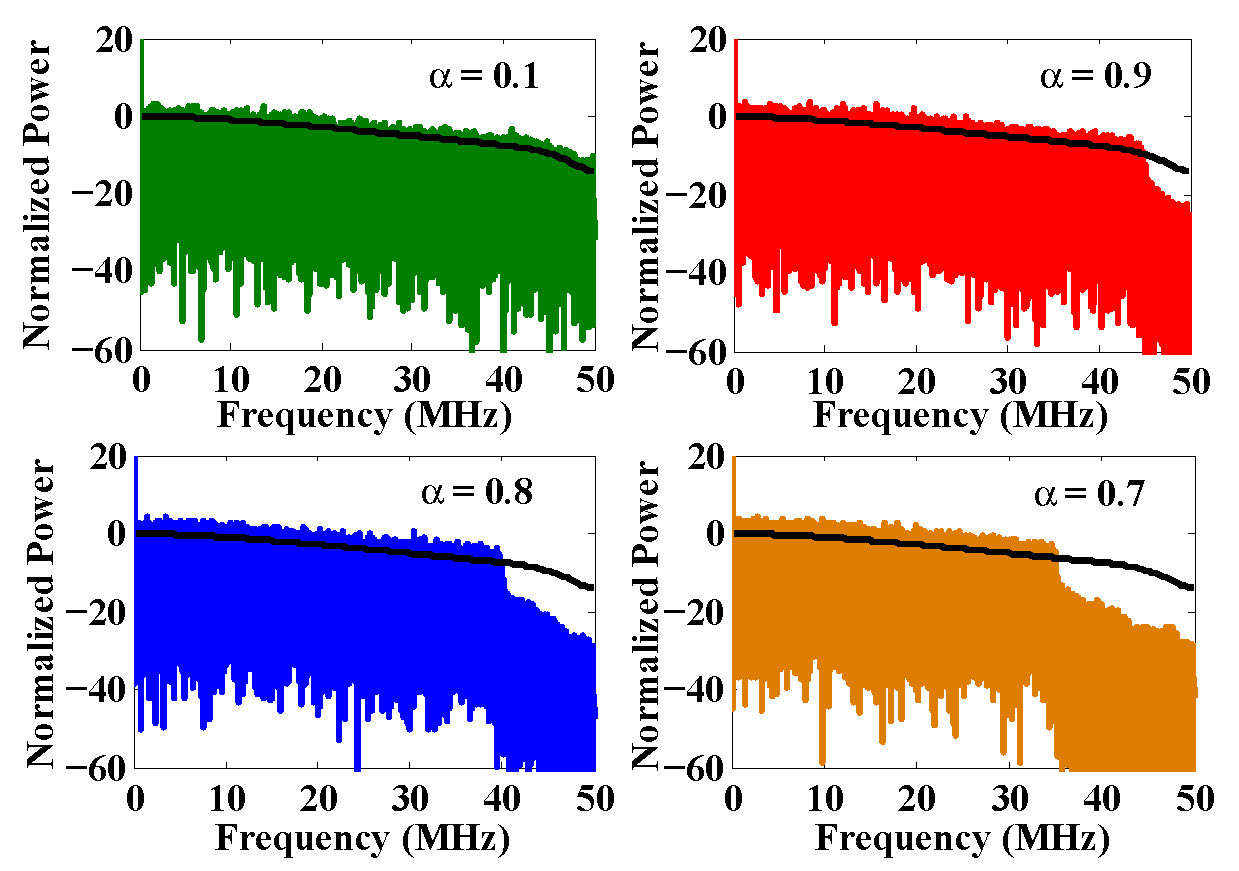}
\caption{The spectra for DCT-OFDM (i.e. $\alpha = 1$) and FrCT-NOFDM signals with transmission rate of 100 Mbit/s after the optical-wireless channel. The dark line denotes the transfer function of the CB model with 3-ns RMS delay spread.}\label{fig:5}
\end{figure}

\section{Simulation Setup and Results} \label{section_5}
\subsection{Simulation Setup}
In this section, we present the simulation setup and the detail of channel equalization and inter-carrier interference (ICI) cancellation algorithm.

For VLC systems, the simulation of FrCT-NOFDM is implemented based on the block diagram shown in Fig. \ref{fig:1}. The encoding block mainly consists of the real constellation mapper, 256-point IFrCT, and cyclic prefix (CP) addition. In our simulation, the modulated constellation employs the $2$-PAM. The CP is $1/16$ of the symbol duration, in which 16 samples are used. In one frame, 256 FrCT-NOFDM symbols and 10 training symbols are transmitted. Eight frames are used to calculate the BER. A suitable DC bias is required to make the FrCT-NOFDM signal to be unipolar. The transmission rate of the generated FrCT-NOFDM signal is set to 100 Mbit/s.

As shown in Section \ref{section_4}, the optical-wireless channel employs the CB model and the adding noise is Gaussian and signal-independent in the simulation. Fig. \ref{fig:5} shows the spectra for the DCT-OFDM (i.e. $\alpha = 1$) and FrCT-NOFDM signals with the transmission rate of $100$ Mbit/s after the optical-wireless channel. The dark line denotes the transfer function of CB model with 3-ns RMS delay spread. After the optical-wireless channel, the signals suffer the high-frequency distortion. The bandwidth of DCT-OFDM signal is equal to $50$ MHz, which is half of the transmission rate. The bandwidth of FrCT-NOFDM signal is smaller than that of DCT-OFDM, which is 45 MHz, 40 MHz, and 35 MHz when the $\alpha$ is set to $0.9$, $0.8$, and $0.7$, respectively. Therefore, FrCT-NOFDM occupies smaller bandwidth and thus achieves higher spectral efficiency compared to DCT-OFDM.

The decoding block mainly consists of CP removal, channel equalization, 256-point FrCT, iterative detection (ID) algorithm, and constellation demapper. The channel estimation and frequency-domain equalization are implemented to compensate the channel distortion. The non-orthogonal subcarriers give rise to ICI, which seriously deteriorates the BER performance. The ID algorithm can be employed to reduce ICI.

\begin{figure}[!t]
\centering
\includegraphics[width=3in]{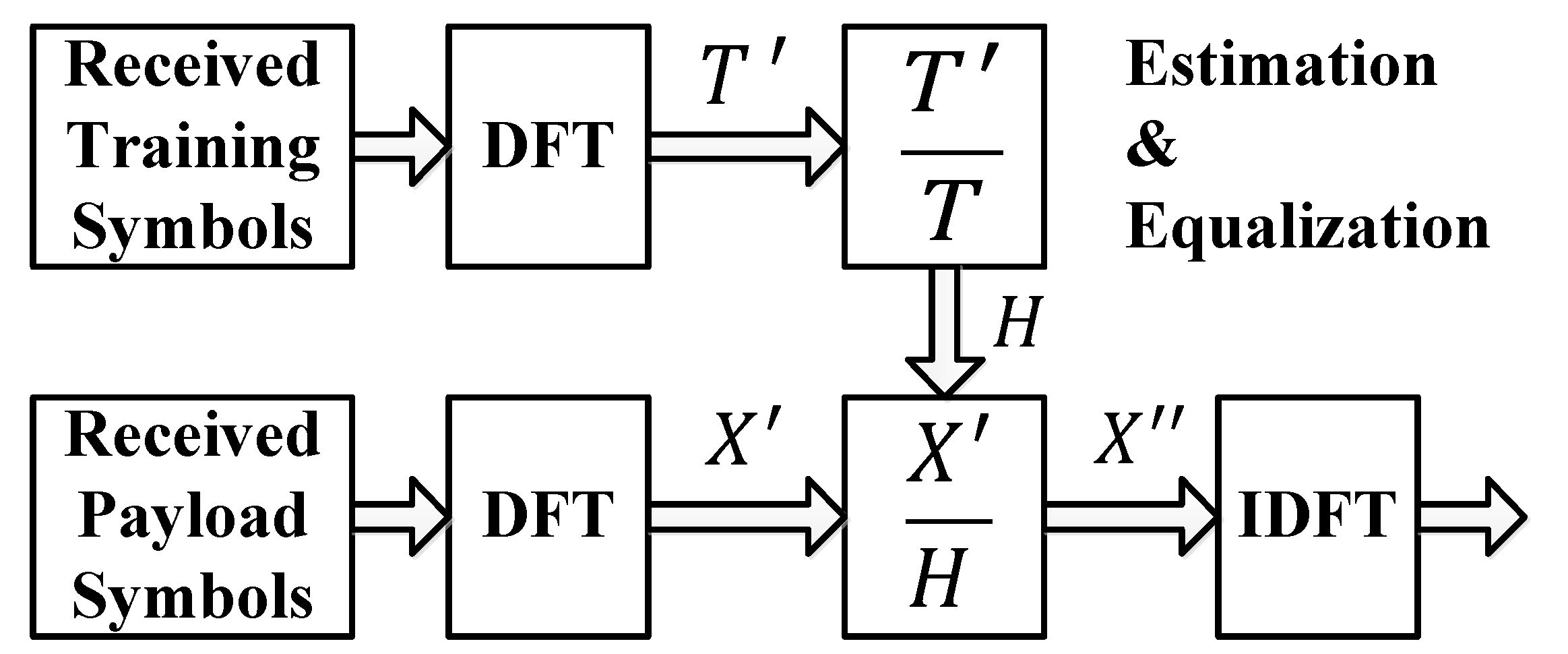}
\caption{Block diagram of channel estimation and frequency-domain equalization for FrCT-NOFDM.}\label{fig:6}
\end{figure}

Figure \ref{fig:6} depicts the block diagram of channel estimation and frequency-domain equalization for FrCT-NOFDM. In the channel estimation, the training symbols are employed to estimate the channel characteristic. The received training symbols are sent to the DFT module to obtain its frequency-domain symbols $\emph{\textbf{T}}'$. The channel matrix $\emph{\textbf{H}}$ can be calculated by $\emph{\textbf{T}}'/\emph{\textbf{T}}$ where $\emph{\textbf{T}}$ are the transmitted training symbols in the frequency domain. In the frequency-domain equalization, the received payload symbols are firstly transformed from the time domain to the frequency domain. The frequency-domain payload symbols $\emph{\textbf{X}}'$ are divided by the channel matrix $\emph{\textbf{H}}$ to realize the equalization. After equalization, the output symbols $\emph{\textbf{X}}''$ are sent to IDFT module. The outputs of IDFT module are the FrCT-NOFDM symbols after equalization.

\renewcommand{\algorithmicrequire}{ \textbf{Input:}}
\renewcommand{\algorithmicensure}{ \textbf{Output:}}
\begin{algorithm}[!t]
\caption{ID algorithm for FrCT-NOFDM.}
\label{alg:1}
\begin{algorithmic}[1]
\Require Received symbol : $\textbf{R}$; Correlation matrix : $\textbf{C}$; Iterative number : $I$; Identity matrix : $\textbf{e}$.
\Ensure Recovered symbol : $\textbf{S}$.
\State Initialization : $\textbf{S}_0 = 0$, $d = 1$ \Comment{Beginning of iteration}
\For{$i = 1$; $i\leq I$; $i++$}
\State $\textbf{S}_i = \textbf{R} - (\textbf{C}-\textbf{e})\textbf{S}_{i-1}$  \Comment{Eliminating the ICI}
\If{$\textbf{S}_i > d$} \Comment{Signal decision}
\State $\textbf{S}_i = 1$
\ElsIf{$\textbf{S}_i < -d$}
\State $\textbf{S}_i = -1$
\Else
\State $\textbf{S}_i = \textbf{S}_i$
\EndIf
\State $d = 1-i/I$ \Comment{Updating $d$}
\EndFor
\State $\textbf{S} = \textbf{S}_I$
\State Return $\textbf{S}$
\end{algorithmic}
\end{algorithm}

Algorithm \ref{alg:1} depicts the detailed processing of the ID algorithm for FrCT-NOFDM with $2$-PAM constellation. The correlation matrix $\textbf{C}$ can be calculated by,
\begin{equation}\label{eq:eq19}
\begin{aligned}
C_{l,~m} &=\frac{2}{N}\sum\limits_{k=0}^{N-1}W_{l}\text{cos}(\frac{\alpha \pi l(2k+1)}{2N})\\
&~~~~~~~~~~~~~~~~\times W_{m}\text{cos}(\frac{\alpha \pi (2k+1)m}{2N}),
\end{aligned}
\end{equation}
in which the elements $C_{l,m}$ are the values of cross-correlation representing the interference between subcarriers $l$ and $m$.

The ID algorithm can be implemented by three steps. The first step reduces the ICI by
\begin{equation}\label{eq:eq20}
\textbf{S}_i = \textbf{R} - (\textbf{C}-\textbf{e})\textbf{S}_{i-1}
\end{equation}
where $i$ denotes the $i$-th iteration. The second step is the decision operation. The signals falling on the decision regions (i.e., the value of the signal is larger than $d$ or smaller than $-d$) can be mapped to the corresponding constellation points. The last step is updating the decision level by $d = 1-i/I$ where $I$ is the total iterative number. Finally, the recovered symbol $\textbf{S}$ can be obtained after the $I$-th iteration.

\subsection{Simulation Results}
In this section, we give the simulation results of FrCT-NOFDM for the VLC systems. The BER performance of FrCT-NOFDM is comprehensively analyzed under different $\alpha$, RMS delay spread, DC bias, and iterative number. Meanwhile, we demonstrate that FrCT-NOFDM has the superior security performance.

\begin{figure}[!t]
\centering
\includegraphics[width=3in]{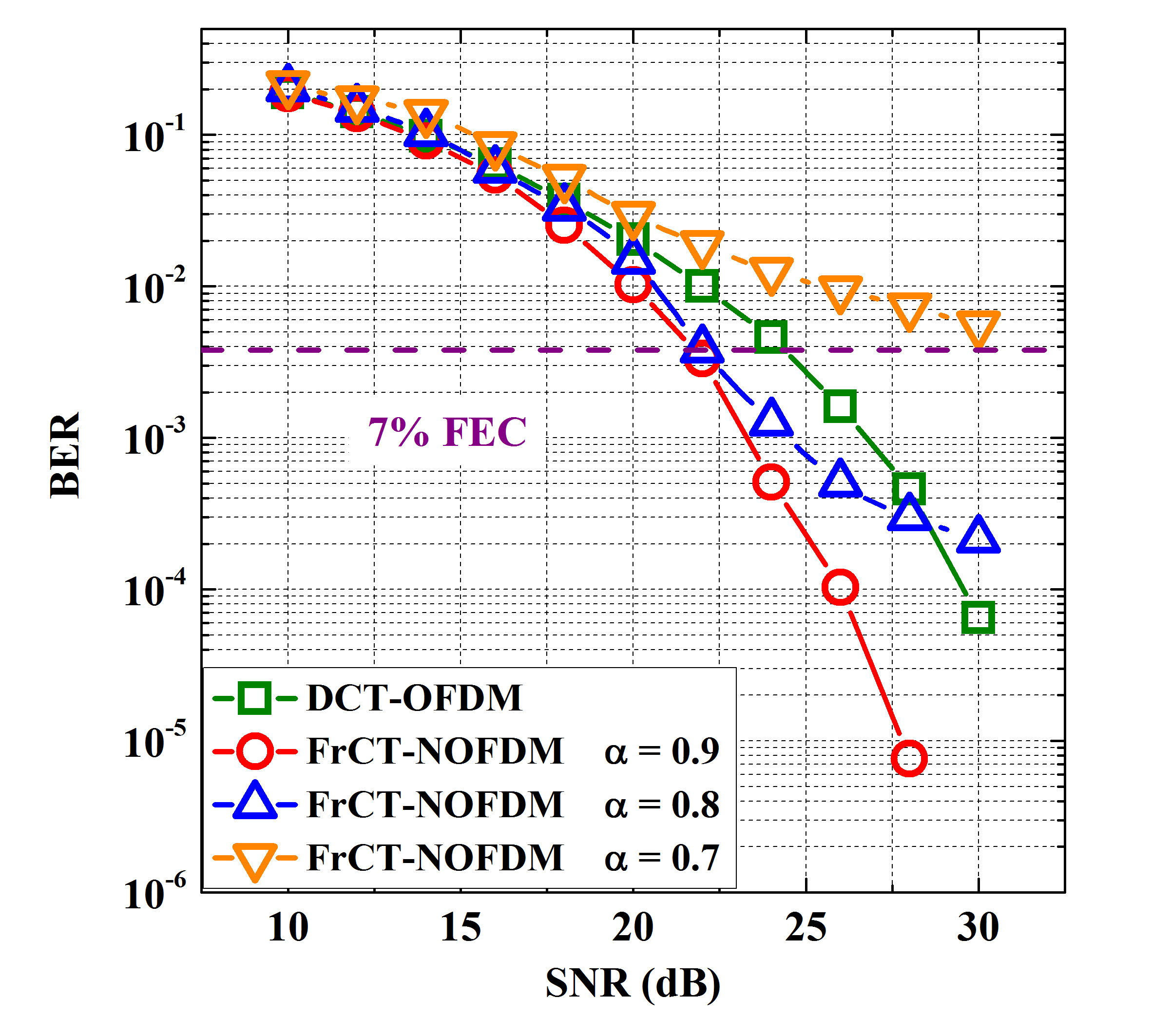}
\caption{BER versus the SNR for DCT-OFDM and FrCT-NOFDM signals.}\label{fig:7}
\end{figure}

\begin{figure}[!t]
\centering
\includegraphics[width=3in]{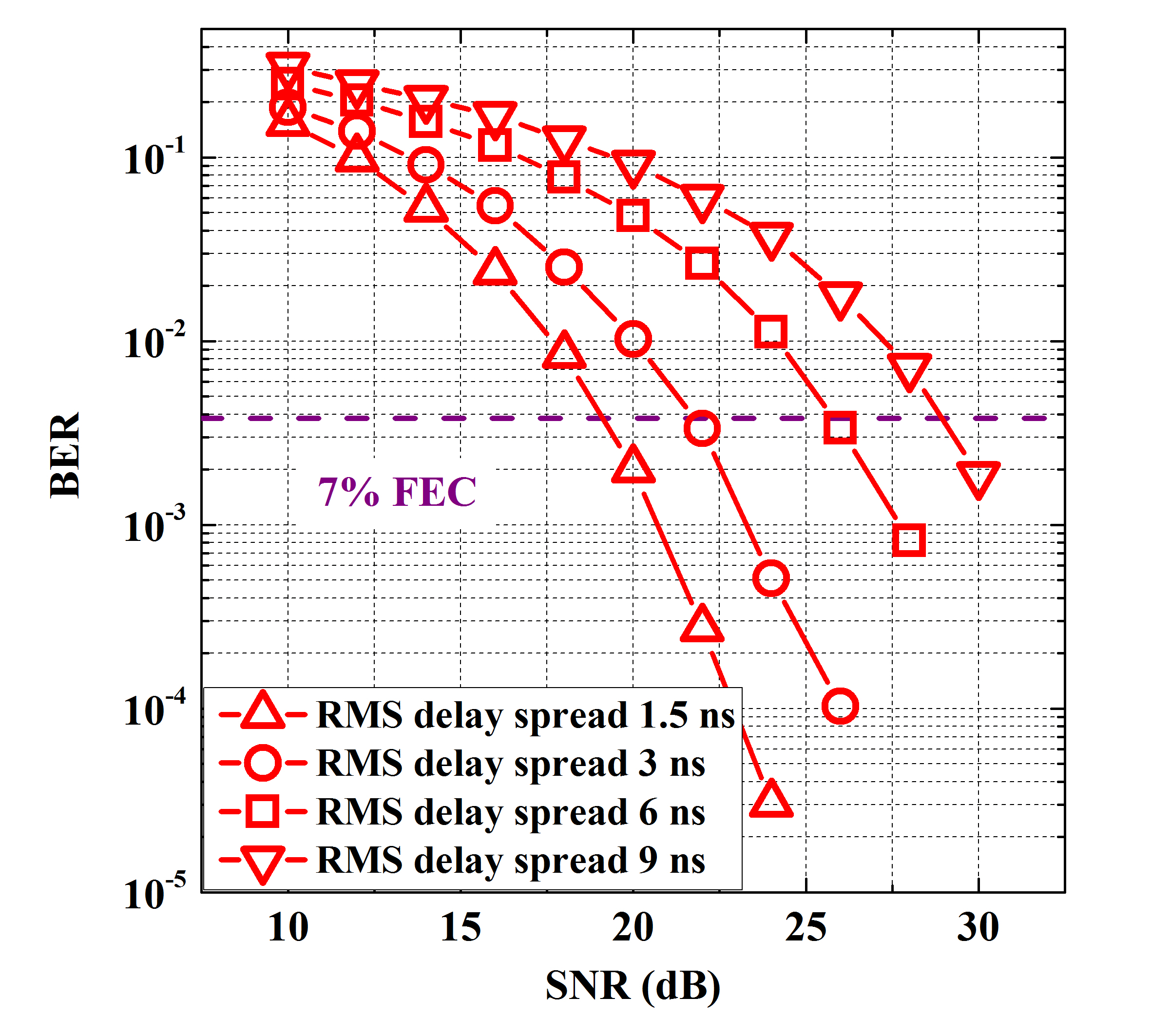}
\caption{BER against the SNR for different RMS delay spread of the optical-wireless channel.}\label{fig:8}
\end{figure}

Figure \ref{fig:7} shows BER versus the SNR for DCT-OFDM and FrCT-NOFDM signals. The RMS delay spread is set to $3$ ns, the DC bias is set to 7 dB, and the iterative number of ID algorithm is set to 20. When the SNR is less than 22 dB, FrCT-NOFDM with $\alpha$ of $0.9$ has almost the same BER performance compared to that with $\alpha$ of $0.8$. Their BER can achieve the $7\%$ forward error correction (FEC) limit at the SNR of 22 dB. In DCT-OFDM, the BER can achieve the $7\%$ FEC limit at the SNR of 24.2 dB. When $\alpha$ is set to $0.9$ or $0.8$, FrCT-NOFDM exhibits an improvement in the SNR of 2.2 dB compared to DCT-OFDM. As shown in Fig. \ref{fig:5}, the power of high-frequency part in DCT-OFDM signal is seriously declined, but the high-frequency part in FrCT-NOFDM is empty due to the compression of bandwidth. Therefore, DCT-OFDM suffers more high-frequency distortion and has worse BER performance compared to FrCT-NOFDM with $\alpha$ of $0.9$ and $0.8$.

After ID algorithm, the residual ICI in FrCT-NOFDM increases with the decreasing of $\alpha$. As Fig. \ref{fig:7} depicts, when the SNR is larger than $22$ dB, the BER of FrCT-NOFDM with $\alpha$ of $0.8$ is higher than that of FrCT-NOFDM with $\alpha$ of 0.9. This is because the residual ICI in FrCT-NOFDM with $\alpha$ of $0.8$ is larger than that in FrCT-NOFDM with $\alpha$ of $0.9$, and with the increase of SNR, the residual ICI turns into the major distortion. Due to the large residual ICI, FrCT-NOFDM with $\alpha$ of $0.7$ can not achieve the $7\%$ FEC limit at the SNR of $30$ dB.

Figure \ref{fig:8} shows the BER against the SNR for different RMS delay spread of the optical-wireless channel. The $\alpha$ of FrCT-NOFDM is set to $0.9$, the DC bias is set to $7$ dB and the iterative number of ID algorithm is set to 20. The BER performance deteriorates with the increase of the RMS delay spread due to the decrease of the channel bandwidth. When RMS delay spread is set to $9$ ns, the bandwidth of the CB model is only $8.3$ MHz, which is insufficient for the signal with $45$-MHz bandwidth. Under this condition, BER can only achieve the $7\%$ FEC limit when SNR is $30$ dB.

\begin{figure}[!t]
\centering
\includegraphics[width=3in]{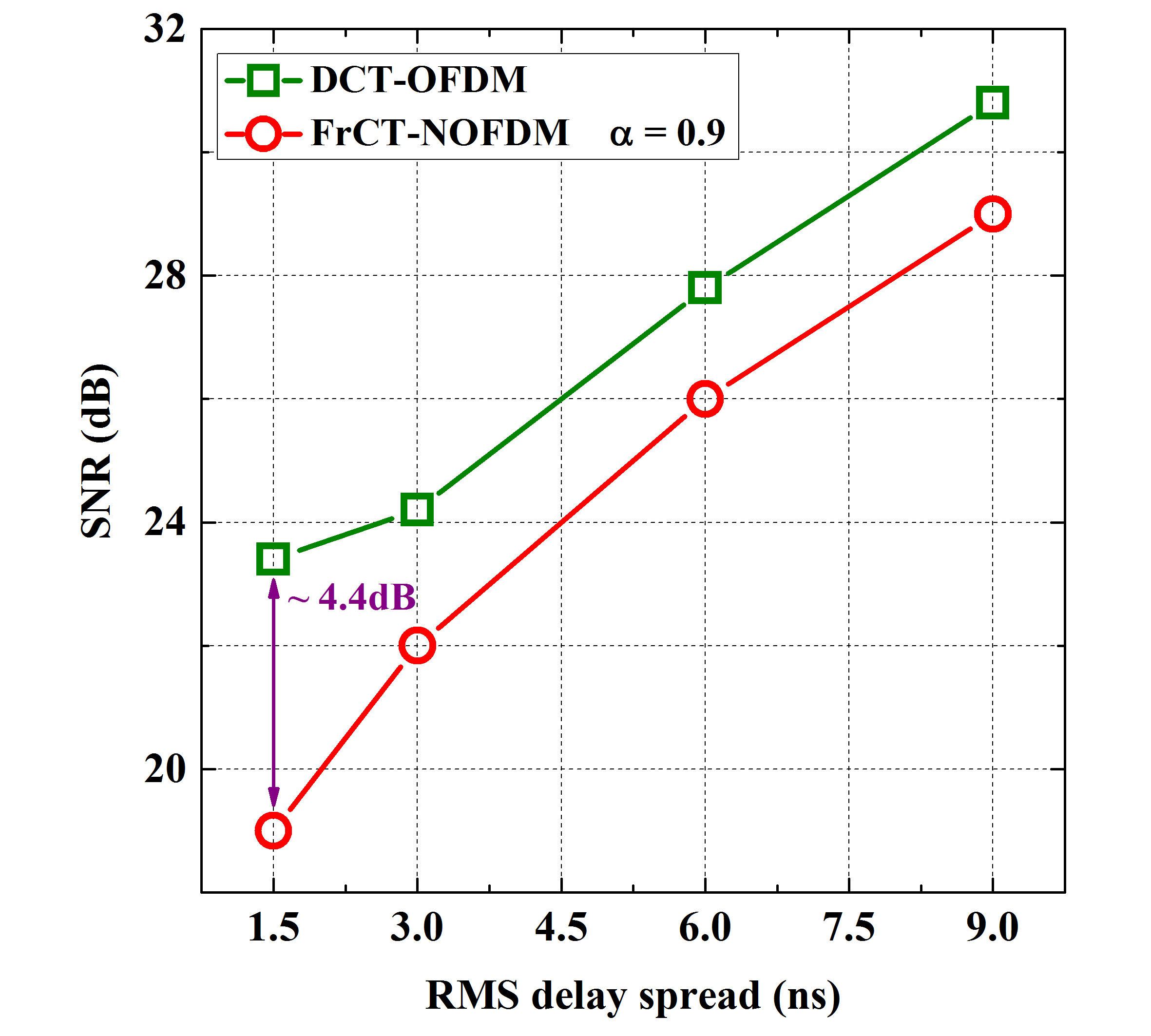}
\caption{The required SNR at the $7\%$ FEC limit against the RMS delay spread for DCT-OFDM signal and FrCT-NOFDM signal with $\alpha$ of 0.9.}\label{fig:9}
\end{figure}

Figure \ref{fig:9} depicts the required SNR at the $7\%$ FEC limit against the RMS delay spread for DCT-OFDM and FrCT-NOFDM with $\alpha$ of $0.9$. When RMS delay spread is set to $1.5$ ns, the required SNR for DCT-OFDM signal is about $4.4$-dB higher than that for FrCT-NOFDM signal. As shown in Fig. \ref{fig:4}, when the RMS delay spread is set to $1.5$ ns, the transfer function of the CB model is fading fast while the bandwidth is larger than $45$ MHz, looking like a cliff. This fast high-frequency power fading seriously degrades the performance of the subcarriers between $45$ MHz and $50$ MHz in DCT-OFDM. However, there is no subcarrier between $45$ MHz and $50$ MHz in FrCT-NOFDM with $\alpha$ of $0.9$, thus it is almost not influenced by that fast high-frequency power fading. When RMS delay spread increases, the high-frequency power fading becomes smooth. Therefore, the difference of the required SNR between DCT-OFDM and FrCT-NOFDM is no longer so large. The required SNR in FrCT-NOFDM is approximately $2$ dB smaller than that in DCT-OFDM when RMS delay spread is equal or greater than $3$ ns.

\begin{figure}[!t]
\centering
\includegraphics[width=3in]{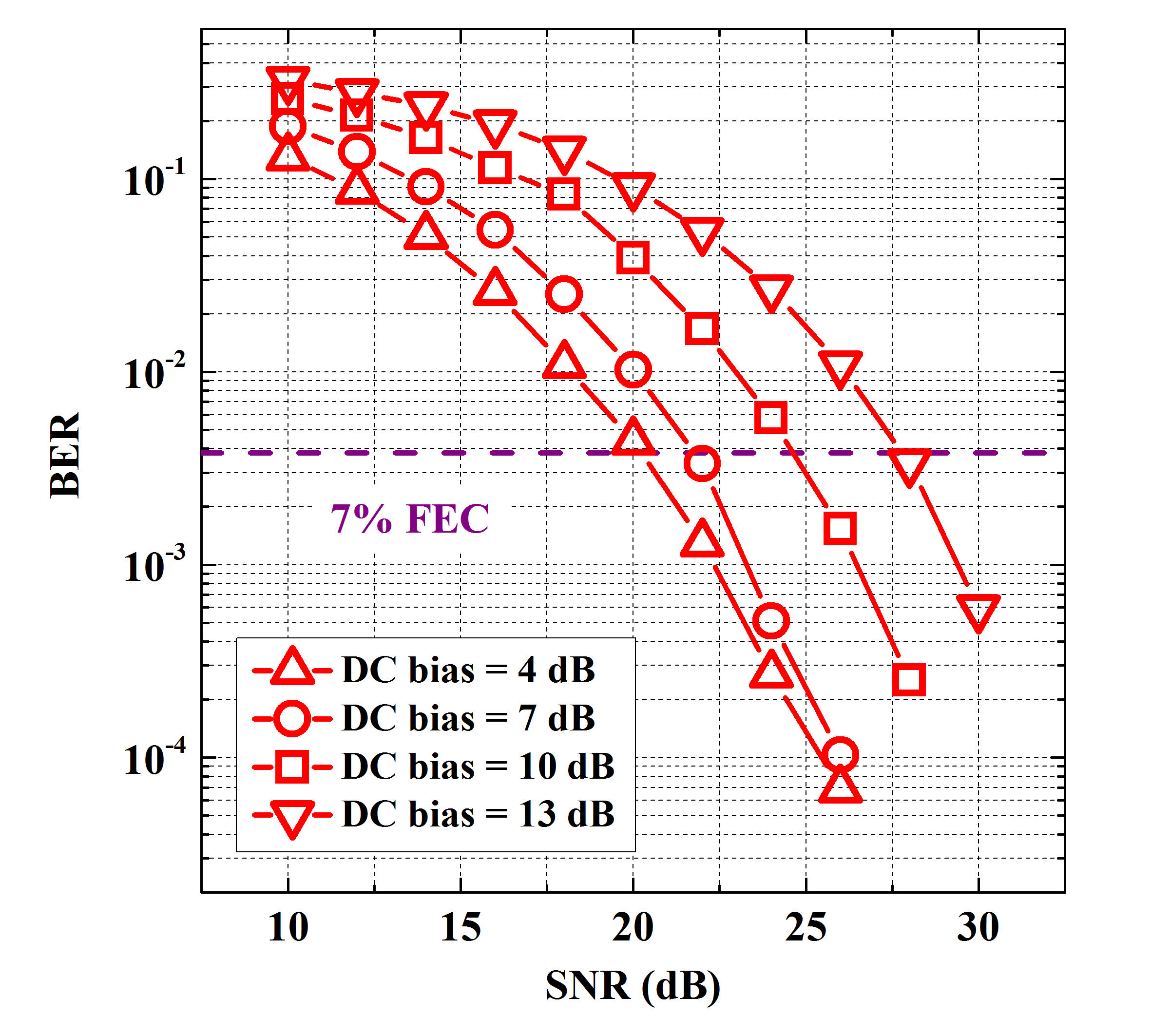}
\caption{BER versus the SNR for different DC bias in FrCT-NOFDM signal when $\alpha$ is set to $0.9$.}\label{fig:10}
\end{figure}

\begin{figure}[!t]
\centering
\includegraphics[width=3in]{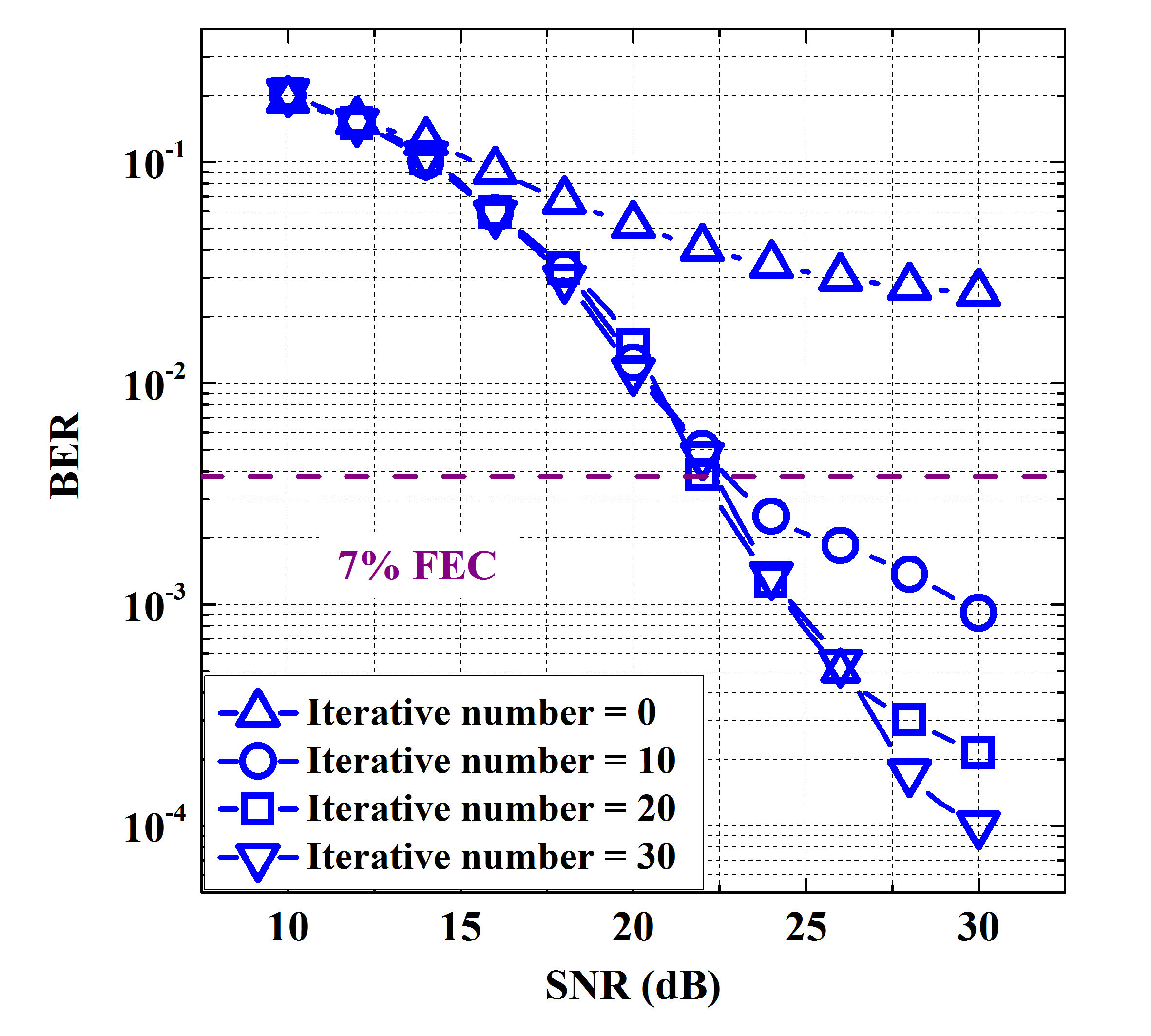}
\caption{BER against the SNR for different iterative numbers of the ID algorithm when the $\alpha$ is set to $0.8$ in FrCT-NOFDM.}\label{fig:11}
\end{figure}

Figure \ref{fig:10} shows the BER versus the SNR for different DC bias in FrCT-NOFDM when $\alpha$ is set to $0.9$. The iterative number of ID algorithm is set to $20$ and the RMS delay spread is set to $3$ ns. As shown in Eq. (\ref{eq:eq13}), FrCT-NOFDM suffers the clipping distortion, which decreases with the increase of DC bias. If the DC bias is large enough, there is almost no clipping noise in FrCT-NOFDM and its power is approximately equal to the power of useful signal plus the power of DC bias as shown in Eq. (\ref{eq:eq14}). Therefore, when the DC bias is large enough, the difference between the required SNRs for FrCT-NOFDM with different DC bias is equal to the difference between the corresponding DC-bias power. When the DC bias is $4$ dB, the signal still suffers the clipping noise, thus the difference between the required SNRs of FrCT-NOFDM with $4$- and $7$-dB DC bias is smaller than $3$ dB. However, when DC bias is larger than 7 dB, there is little clipping noise in the signal. The difference between the required SNRs of FrCT-NOFDM with $7$- and $10$-dB DC bias is almost equal to $3$ dB. The simulation result coincides with the theoretical analysis.

Figure \ref{fig:11} shows the BER against the SNR for different iterative numbers of the ID algorithm. The $\alpha$ of FrCT-NOFDM is set to $0.8$, the RMS delay spread is set to $3$ ns, and the DC bias is set to $7$ dB. When the ID algorithm is not employed (i.e., the iterative number is set to $0$), the BER cannot achieve the $7\%$ FEC limit although the SNR is set to $30$ dB. This is because the ICI seriously degrades the BER performance of FrCT-NOFDM. With the increase of the iterative number, the BER performance is markedly improved. Therefore, ID algorithm can effectively eliminate the ICI for FrCT-NOFDM. However, the complexity of ID algorithm increases with the iterative number. The performance of ID algorithm will no longer be obviously improved with the increase of the iterative number while the iterative number is large. Therefore, we can choose the suitable iterative number by synthetically considering the effect and complexity of ID algorithm.

\begin{figure}[!t]
\centering
\includegraphics[width=3in]{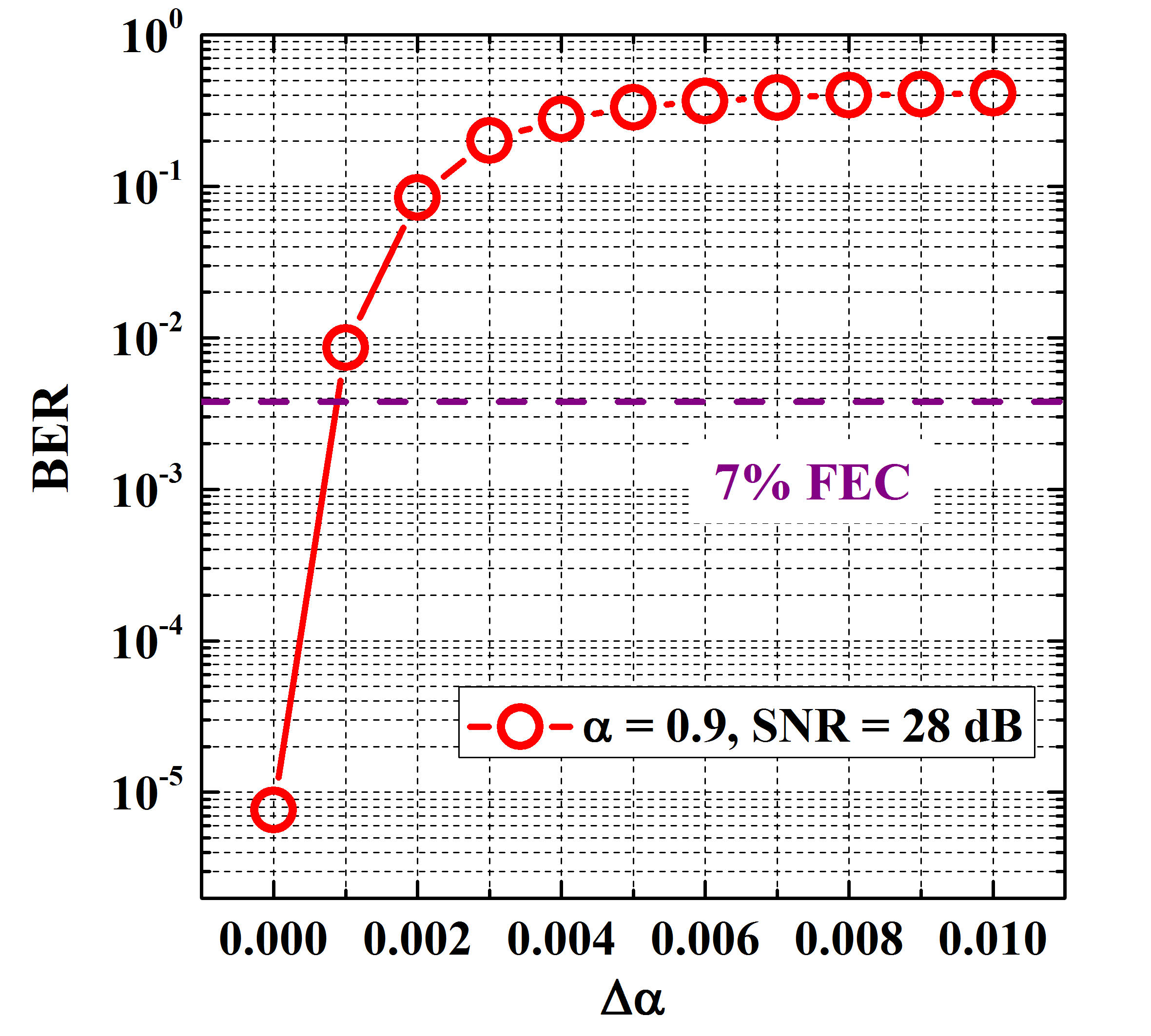}
\caption{BER versus $\Delta \alpha$ for FrCT-NOFDM signal when $\alpha$ at the transmitter is set to $0.9$ and SNR is set to $28$ dB.}\label{fig:12}
\end{figure}

Figure \ref{fig:12} shows the BER versus $\Delta \alpha$ for FrCT-NOFDM signal when $\alpha$ at the transmitter is set to $0.9$ and SNR is set to $28$ dB. The iterative number of ID algorithm is set to $20$ and the RMS delay spread is set to $3$ ns. The $\Delta \alpha$ denotes the deviation of $\alpha$ between the transmitter and receiver. When $\Delta \alpha$ is larger than $0.001$, the BER performance will be significantly deteriorated (BER $>$ $7\%$ FEC limit). In other word, the data can only be recovered accurately when $\Delta \alpha$ is less than $0.001$. This is because both FrCT and ID algorithm at the receiver need an accurate $\alpha$. Therefore, $\alpha$ can be used as an encryption key for the security communications. If the $\alpha$ can not be known accurately, the transmitted data cannot be accurately recovered at the receiver. It reveals that FrCT-NOFDM signal has the superior security performance for application in the security VLC systems.

\section{Conclusion} \label{section_6}
This paper proposed the FTN FrCT-NOFDM signal for VLC systems. Compared to FrFT-NOFDM signal, FrCT-NOFDM signal is real-valued, which can be directly applied to the VLC systems without upconversion. Therefore, FrCT-NOFDM signal is more suitable for cost-sensitive VLC systems. Under the same transmission rate, FrCT-NOFDM signal occupies smaller bandwidth compared to OFDM signal. When $\alpha$ is set to $0.8$, $20\%$ bandwidth saving can be obtained. By this way, FrCT-NOFDM signal suffers less high-frequency distortion, which is suited to the bandwidth-limited VLC systems.

We presented the simulations to investigate the performance of FrCT-NOFDM. When the RMS delay spread is set to $3$ ns (i.e., the 3-dB channel bandwidth is $20.9$ MHz) and the transmission rate is set to 100 Mbit/s, FrCT-NOFDM with $\alpha$ of $0.9$ or $0.8$ exhibits an improvement in the SNR of 2.2 dB compared to DCT-OFDM because of the less high-frequency distortion. Meanwhile, FrCT-NOFDM has the superior security performance for application in the security VLC systems. However, it is worth noting that the residual ICI after ID algorithm is still a critical problem for the high-order constellations and thus the more effective algorithm will be investigated to solve this problem in our future work. In conclusion, FrCT-NOFDM shows the potential for application in the future VLC systems.


%





\ifCLASSOPTIONcaptionsoff
  \newpage
\fi

\end{document}